\documentclass{ws-procs9x6}
\begin{document}
\title{SCALING OF ENTANGLEMENT ENTROPY AND HYPERBOLIC GEOMETRY}
\author{HIROAKI MATSUEDA}
\address{Advanced course of information and electronic system engineering, Sendai National College of Technology,\\
Sendai, Miyagi 989-3128, Japan\\
$^*$E-mail: matsueda@sendai-nct.ac.jp}
\begin{abstract}
Various scaling relations of the entanglement entropy are reviewed. Based on the scaling, I would like to point out similarity of mathematical formulation among recent topics in wide research area. In particular, the scaling plays crucial roles on identifying a quantum system with a physically different classical system. Close connection between the scaling and hyperbolic geometry and contrast between bulk/edge correspondence and compactification for the identification are also addressed.
\end{abstract}
\keywords{Entanglement Entropy; Singular Value Decomposition (SVD); Area Law; Black Hole Thermodynamics; Density Matrix Renormalization Group (DMRG); Matrix Product State (MPS); Tensor Product State (TPS) / Projected Entangled Pair State (PEPS); Multiscale Entanglement Renormalization Ansatz (MERA); Hyperbolic Geometry;  Anti-de Sitter Space / Conformal Field Theory (AdS/CFT) correspondence ; Compactification ; Quantum Monte Carlo Simulation (QMC) ; Image Processing.}
\bodymatter

\section{Introduction}

The entanglement entropy is one of the most fundamental concepts in quantum information. Recently, its efficiency as a tool to see through underlying physical principles in our targets is also recognized in statistical physics, condensed matter physics, and string theory. This wide applicability comes from universality of the entropy irrespective of their details. Roughly speaking, the entropy represents a logarithm of a correlation function. Thus, the entropy directly picks up critical exponents. In addition, the entropy can detect topological structure of the manifold where the target model is defined. In order to evaluate the universal feature quantitatively, it is necessary to find scaling relations of the entropy as a function of the linear system size. Furthermore, the scaling relations tell us how physically different systems are associated with each other. When two systems have the same scaling relation, their eigenvalue spectra of the density matrices may be similar in some cases. The identification between the physically different systems leads to deep understanding of duality, holography, and quantum-classical correspondence. They are particularly important ideas in string theory and Quantum Monte Carlo simulation. In ergodic theory, it was a central problem to examine what is a class of measure-preserving transformations in which the information entropy determines isomorphic properties. In viewpoints of symmetry and group theory, the scaling, conformal invariance, and hyperbolic geometry are mutually correlated. Therefore, a lot of important concepts of physics, mathematics, and information merge together in the examination of the entropy. In this article, I would like to review some aspects related to the entropy scaling and its underlying geometrical structure.

\section{Entanglement Entropy and Singular Value Decomposition}

Let us consider a spatially $d$-dimensional ($d$D) quantum system that is composed of a subsystem $A$ and an environment $B$. The system is sometimes called 'universe' or 'superblock'. The linear size of $A$ is denoted by $L$. The entanglement entropy represents the amount of information across the boundary between $A$ and $B$. We start with a pure state of the universe
\begin{eqnarray}
\left|\psi\right>=\sum_{x,y}\psi(x,y)\left|x\right>\left|y\right>,
\end{eqnarray}
where $\left|x\right>$ and $\left|y\right>$ represent basis states of $A$ and $B$, respectively. The density matrix of the subsystem $A$ is then defined by
\begin{eqnarray}
\rho_{A}=tr_{B}\left|\psi\right>\left<\psi\right|, \label{HM-dm}
\end{eqnarray}
where the symbol $tr_{B}$ traces over degrees of freedom inside of $B$. Then, the entanglement entropy $S_{A}$ is given by
\begin{eqnarray}
S_{A}=-tr_{A}(\rho_{A}\ln\rho_{A}). \label{HM-EE}
\end{eqnarray}
In order to see physical meaning of the entropy, it is better to introduce the singular value decomposition (SVD) of the wave function $\psi(x,y)$. The SVD of $\psi(x,y)$ is defined by
\begin{eqnarray}
\psi(x,y)=\sum_{l}U_{l}(x)\sqrt{\lambda_{l}}V_{l}(y), \label{HM-SVD}
\end{eqnarray}
where $U_{l}(x)$ and $V_{l}(y)$ are the column unitary matrices, and $\lambda_{l}$ is the singular value that is positively definite. In the following, we use the normalized singular value $p_{l}=\lambda_{l}/\sum_{l}\lambda_{l}$. Since $\sum_{l}p_{l}=1$, $p_{l}$ represents a probability of realization of the state labeled by $l$. The entanglement entropy is then expressed as
\begin{eqnarray}
S=S_{A}=S_{B}=-\sum_{l}p_{l}\ln p_{l}. \label{HM-ABequal}
\end{eqnarray}
This simple relation $S_{A}=S_{B}$ clearly shows non-extensivity of the entanglement entropy, since the volume of $A$ is in general different from that of $B$. The object that is common in between is their boundary. Then, the entropy would be proportional to the boundary area. This is called area law scaling. In this sense, the entanglement entropy and the thermal entropy behave quite differently. Therefore, to confirm the area law and to find its violation in specific cases are two active topics.

If we assume equal probability condition $p_{l}=1/m$ for any $l$, the entropy yields the standard Boltzmann's law
\begin{eqnarray}
S=-\sum_{l=1}^{m}\frac{1}{m}\ln\frac{1}{m}=\ln m.
\end{eqnarray}
However, the equal probability condition for the states within the subsystem $A$ is clearly violated in low-dimensional quantum systems. It is easy to see this feature by, for instance, exact diagonalization calculation and other techniques. Thus, the difference between entanglement and thermal entropies is essential. Furthermore, the singular values are uniquely determined after the decomposition, while the column unitary matrices are not. The universal behavior of the entropy is thus due to the presence of a set of the universal singular values.

It is noted that a recent trend to examine the entanglement structure of quantum systems is to introduce the entanglement spectrum as well as the entanglement entropy. The spectrum is defined by
\begin{eqnarray}
E_{l}=-\ln\lambda_{l}.
\end{eqnarray}
Extensive examinations show that the spectrum is a powerful tool to characterize topological nature of the system. The nature is govened by presence or absense of the entanglement gap. The reason for the powerfulness comes from a fact that a topological phase is a phase of matter that cannot be described by an order parameter.

\section{Scaling of Entanglement Entropy}

\subsection{Historical Roots of Anomalous Entropy Scaling}

The thermal entropy is usually an extensive parameter. One exceptional case can be seen in black hole thermodynamics~\cite{HM-Bekenstein,HM-Hawking1,HM-Hawking2}. It is well known that the Bekenstein-Hawking entropy is proportional to the surface area $A=4\pi r^{2}$ of the event horizon of a black hole:
\begin{eqnarray}
dS=\frac{k_{B}c^{3}}{4G\hbar}dA, \label{HM-BH}
\end{eqnarray}
Here, $r=2GM/c^{2}$ is the Schwartzshild's radius. Starting with the surface gravity $\kappa=GM/r^{2}$, the derivative form of $A$ leads to $dM=(\kappa/8\pi G)dA$. Due to the relativistic energy $E=Mc^{2}$ and the first law of thermodynamics $dE=TdS$, we obtain $dS=(\kappa c^{2}/8\pi GT)dA$. If we assume the Hawking temperature $T=\hbar\kappa/2\pi k_{B}c$, the Eq.~(\ref{HM-BH}) is derived. The information absorbed into the black hole can not go out, if we neglect Hawking radiation. Then, a role of the black hole on the entropy is like enviromental degrees of freedom in Eq.~(\ref{HM-dm}), since in the definition of the entanglement entropy an observer in $A$ can not access the information in $B$. Thus, the information theory based on the entanglement entropy has close connection to the black hole physics.

\subsection{General coodinate transformation and horizon}

\begin{figure}
\centerline{\includegraphics[width=8cm]{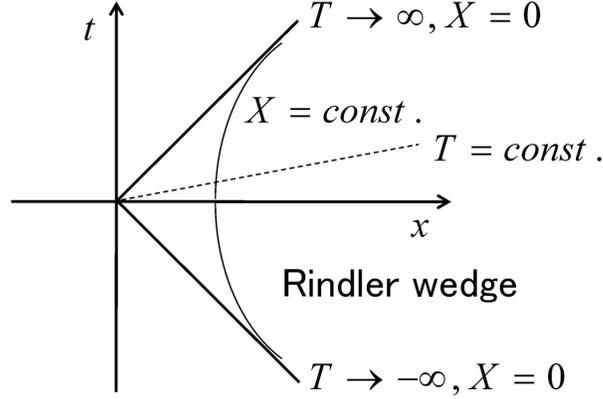}}
\caption{Minkowski and Rindler coodinates.}\label{HM-fig1}
\end{figure}

Consider the metric that does not have any singularities for a static observer. However, an accerated observer may look at different spacetime structure. Actually, the flat Minkowski metric can be transformed into the Rindler one that has event horizon~\cite{HM-Rindler,HM-Davies}. This means that general coodinate transformations beyond Lorentzian ones limits spacetime region that the accerelated observer can access. As shown in Fig.~\ref{HM-fig1}, the Rindler observer is confined in the Rindler 'wedge' which acts as a subsystem $A$, while the outside of the wedge represents the enviromental degrees of freedom $B$. Thus, the thermal entropy for the field propagating in this geometry behaves as the black hole entropy.

Starting with the Minkowski metric (we consider 1D case for simplicity)
\begin{eqnarray}
ds^{2}=-dt^{2}+dx^{2},
\end{eqnarray}
we introduce the following transformation
\begin{eqnarray}
X&=&\sqrt{x^{2}-t^{2}}, \\
T&=&\tanh^{-1}\left(\frac{t}{x}\right),
\end{eqnarray}
or more explicitely
\begin{eqnarray}
t&=&X\sinh T, \\
x&=&X\cosh T,
\end{eqnarray}
where $-\infty<T<\infty$, and $0<X<\infty$. Then, the metric for the new coodinate is given by
\begin{eqnarray}
ds^{2}=-X^{2}dT^{2}+dX^{2}.
\end{eqnarray}
We clearly see that the accesible region is only the right Rindler wedge as shown in Fig.~\ref{HM-fig1}.

As a tutorial example, let us consider the massless scalar field $\phi$ propagating in the Rindler space. The action is given by
\begin{eqnarray}
I&=&\int dTdX\sqrt{-g}\left(-\frac{1}{2}g^{\mu\nu}\partial_{\mu}\phi\partial_{\nu}\phi\right).
\end{eqnarray}
The Lagrange equation is obtained as
\begin{eqnarray}
\frac{\partial^{2}}{\partial T^{2}}\phi=\left(X^{2}\frac{\partial^{2}}{\partial X^{2}}+X\frac{\partial}{\partial X}\right)\phi,
\end{eqnarray}
and yields the following solution
\begin{eqnarray}
\phi(T,X)=Ae^{i\Omega(t-x)}=A\exp\left(i\Omega X e^{-T}\right).
\end{eqnarray}
By taking the inverse Fourier transform with respect to $T$, we see that the power spectrum of this field is not monochromatic
\begin{eqnarray}
\int_{-\infty}^{\infty}dT\exp\left(i\Omega X e^{-T}\right)e^{i\omega T}=(-i\Omega X)^{i\omega}\Gamma(-i\omega),
\end{eqnarray}
where $\omega\ne 0$. This leads to Planck distribution with effective inverse temperature $\beta=2\pi$. Complete calculation of the thermal entropy, its correspondence to the Bekenstein-Hawking formula, and related aspects can be seen in the references~\cite{HM-Susskind,HM-Kobat,HM-Emparan,HM-Padmanabhan,HM-Jacobson}.

\subsection{Area Law and its Logarithmic Violation at Criticality}

It has been extensively examined how the entanglement entropy behaves as functions of $L$ and $d$. As already mentioned above, the most well-known formula is the area-law scaling~\cite{HM-Bombelli,HM-Srednicki}
\begin{eqnarray}
S\propto \left(\frac{L}{a}\right)^{d-1},
\end{eqnarray}
which tells us non-extensivity of $S$ in contrast to the thermal entropy. This relation is strictly hold in gapped quantum systems.

The violation of the area law occurs in cases of 1D critical systems~\cite{HM-Holzhey,HM-Calabrese1,HM-Calabrese2,HM-LesHouches}. The entropy is given by
\begin{eqnarray}
S=\frac{1}{3}c\ln\frac{L}{a}, \label{HM-CFT}
\end{eqnarray}
where $c$ is the central charge and $a$ is lattice cutoff. This was first derived from the conformal field theory (CFT), and then numerically confirmed in various models of statistical physics~\cite{HM-Plenio,HM-Riera,HM-Vidal1}. Away from a critical point, the entropy is deformed as
\begin{eqnarray}
S=\frac{1}{6}c{\cal A}\ln\frac{\xi}{a},\label{HM-deformed}
\end{eqnarray}
with correlation length $\xi$ and the number of boundary points ${\cal A}$ of $A$.

\subsection{Fermi Surfaces and Entropy}

The violation also occurs in models with the Fermi surfaces~\cite{HM-Wolf,HM-Gioev1,HM-Gioev2,HM-Li,HM-Barthel}. The Fermi surface devides the momentum space into two sectors: sets of occupied and unoccupied states. The electronic excitations normal to the Fermi surface are like  those of chiral Luttinger liquid. The excitations lead to logarithmic correction. The entropy is then given by
\begin{eqnarray}
S&=&\frac{1}{3}C\left(\frac{L}{a}\right)^{d-1}\log\frac{L}{a} +B\left(\frac{L}{a}\right)^{d-1}+A\left(\frac{L}{a}\right)^{d-2}+\cdots, \\
C&=&\frac{1}{4(2\pi)^{d-1}}\int_{\partial\Omega}\int_{\partial\Gamma}\left|n_{x}\cdot n_{p}\right|dA_{x}dA_{p},
\end{eqnarray}
where $C$ is the number of excitation modes across the Fermi surface $\partial\Gamma$, $\partial\Omega$ is the spatial region considered, and $n_{x}$ and $n_{p}$ are the unit normals to the boundaries.

\subsection{Topological Entanglement Entropy}

The physical characterization of topological orders of 2D is a hot topic in condensed matter physics. Usually, the topological nature can be seen as edge excitations and quasiparticle statistics. However, the topological order is manifest in the basic entanglement of the ground state wave function~\cite{HM-Kitaev,HM-Levin}. In $Z_{q}$ lattice gauge theory, the entropy has a subleading term called topological entropy in addition to the area-law scaling
\begin{eqnarray}
S=\alpha L - \ln\sqrt{q} + \cdots.
\end{eqnarray}
The string-net condensed model and the Kitaev's toric code are two important examples of realizing the presence of the topological entropy.

\subsection{Entanglement Support of Matrix Product State}

\begin{figure}
\centerline{\includegraphics[width=6cm]{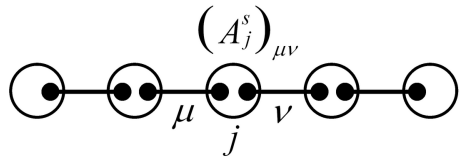}}
\caption{MPS on 1D chain. Each entangled bond is characterized by a symbol $\mu=1,2,...,\chi$. The index $s$ is local degree of freedom ($s=\uparrow,\downarrow$ in $S=1/2$ spin systems).}\label{HM-fig2}
\end{figure}

In statistical physics, there is another type of entropy scaling originated from a recently developed variational approach to quantum systems. The approach optimizes the following matrix product state (MPS) ansatz~\cite{HM-Ostlund,HM-Rommer}
\begin{eqnarray}
\left|\psi\right>=\sum_{\{s_{j}\}}tr(A_{1}^{s_{1}}A_{2}^{s_{2}}\cdots A_{n}^{s_{n}})\left|s_{1}s_{2}\cdots s_{n}\right>, \label{HM-mpswf}
\end{eqnarray}
where the site-dependent matrices $A_{j}^{s_{j}}$ have $\chi\times\chi$ dimensions, and $s_{j}$ represents local degrees of freedom (see Fig.~\ref{HM-fig2}). Historically, matrix product forms play crucial roles on exacly solvable statistical models.

Let us consider the simplest case. We are going to expresse the singlet state $\left|\psi\right>=\left|\uparrow\downarrow\right>-\left|\downarrow\uparrow\right>$ by a particular MPS. It is clear that the direct product state (local decomposition) is not exact
\begin{eqnarray}
\left|\phi\right>&=&\sum_{s_{1}=\uparrow,\downarrow}c_{1}^{s_{1}}\left|s_{1}\right>\otimes\sum_{s_{2}=\uparrow,\downarrow}c_{2}^{s_{2}}\left|s_{2}\right> \\
&=&c_{1}^{\uparrow}c_{2}^{\uparrow}\left|\uparrow\uparrow\right>+c_{1}^{\uparrow}c_{2}^{\downarrow}\left|\uparrow\downarrow\right>+c_{1}^{\downarrow}c_{2}^{\uparrow}\left|\downarrow\uparrow\right>+c_{1}^{\downarrow}c_{2}^{\downarrow}\left|\downarrow\downarrow\right>.
\end{eqnarray}
Actually, we can not simultaneously take $c_{1}^{\uparrow}c_{2}^{\uparrow}=1$, $c_{1}^{\uparrow}c_{2}^{\uparrow}=0$, $c_{1}^{\uparrow}c_{2}^{\downarrow}=0$, and $c_{1}^{\downarrow}c_{2}^{\downarrow}=-1$. However, we can introduce the following expression
\begin{eqnarray}
\left|\psi\right>=\sum_{s_{1},s_{2}}A^{s_{1}}B^{s_{2}}\left|s_{1}s_{2}\right>
\end{eqnarray}
where the two local vectors $A^{s_{1}}$ and $B^{s_{2}}$ are taken to be
\begin{eqnarray}
A^{\uparrow}=\left(
\begin{array}{cc}
x & y
\end{array}
\right), 
A^{\downarrow}=\left(
\begin{array}{cc}
z & w
\end{array}
\right),
\end{eqnarray}
\begin{eqnarray}
B^{\uparrow}=\frac{1}{xw-yz}\left(
\begin{array}{c}
y \cr -x
\end{array}
\right), 
B^{\downarrow}=\frac{1}{xw-yz}\left(
\begin{array}{c}
w \cr -z
\end{array}
\right),
\end{eqnarray}
for scalar variables $x$, $y$, $z$, and $w$. This expression means that the matrix dimension $\chi$ provides quantum correlation between neighboring spins.

The MPS state is known to be exact for gapped 1D systems such as the valence bond solid (VBS) state. Furthermore, the MPS is foundation of the density matrix renormalization group (DMRG) method that is a powerful numerical technique for 1D quantum systems~\cite{HM-White1,HM-White2}. DMRG devides the system into two parts, and then truncates the density matrix eigenstates with the small eigenvalues. Therefore, DMRG is based on entanglement control. The restriction of the matrix dimension upto $\chi$ means that the amount of information described by this MPS is given by
\begin{eqnarray}
S_{\chi}=-\sum_{l=1}^{\chi}\lambda_{l}\ln\lambda_{l},
\end{eqnarray}
where $\lambda_{1}>\lambda_{2}>\cdots >\lambda_{\chi}$. If we assue that $A_{j}^{s_{j}}$ is a scalar variable ($\chi=1$), Eq~(\ref{HM-mpswf}) gives a local approximation as I have already mentioned. Taking a sufficiently large $\chi$ value gives us an asymptotically exact wave function. Thus, the matrix dimension $\chi$ represents how precisely we can take long-range quantum correlation. When we apply this variational method to 1D critical systems, the half-chain entropy (${\cal A}=1$) behaves as~\cite{HM-Tagliacozzo,HM-Pollman,HM-Huang}
\begin{eqnarray}
S_{\chi}=\frac{c\kappa}{6}\ln\chi, \label{HM-mps}
\end{eqnarray}
where the exponent $\kappa$ is defined by
\begin{eqnarray}
\kappa=\frac{6}{c\left(\sqrt{12/c}+1\right)}.
\end{eqnarray}
Comparing Eq.~(\ref{HM-mps}) with Eq.~(\ref{HM-deformed}), we know that
\begin{eqnarray}
\xi=\chi^{\kappa},
\end{eqnarray}
which means that $\chi$ is actually related to the length scale $\xi$. For instance, in the Ising universality class $c=1/2$
\begin{eqnarray}
\kappa&\sim& 2, \\
S_{\chi}&\sim&\frac{1}{6}\ln\chi.
\end{eqnarray}
I have found an old DMRG paper that addresses the correlation length $\xi$ as a function of the truncation number~\cite{HM-Andersson}. The result for 1D free fermion with the central charge $c=1$ is given by
\begin{eqnarray}
\xi=-\frac{1}{\ln\left|1-k\chi^{-\beta}\right|}=\frac{1}{k}\chi^{\beta},
\end{eqnarray}
with $\beta\sim 1.3$ and $k\sim 0.45$. This is really consistent with the above argument where $\kappa\sim 1.33$ for $c=1$.

\section{Holographic Entanglement Entropy: Connection between Scaling and Hyperbolic Geometry}

\subsection{Anti-de Sitter space}

In superstring theory, it has been an interesting topic to find the fundamental mechanism of the anti-de Sitter space (AdS) / CFT correspondence. This is holographic correspondence between a quantum theory, CFT${}_{d+1}$, and general relativity on the AdS space, AdS${}_{d+2}$ (Here, CFT${}_{d+1}$ is abbreviation of spatially $d$-dimensional conformal field theory with one time axis). The AdS space is $(d+2)$-dimensional hyperbolic surface embedded in one higher dimensional flat space $E^{d+1,2}$
\begin{eqnarray}
-X_{-2}^{2}-X_{-1}^{2}+\sum_{k=0}^{d}X_{k}^{2}=-l^{2},
\end{eqnarray}
(see Fig.~\ref{HM-fig3}). This space has two time-like directions $X_{-2}$ and $X_{-1}$. The metric of the AdS space can be represented as
\begin{eqnarray}
ds^{2}&=&g_{\mu\nu}dx^{\mu}dx^{\nu} \\
&=&\frac{l^{2}}{z^{2}}\left(dz^{2}+\eta_{ij}dx^{i}dx^{j}\right), \label{HM-AdS}
\end{eqnarray}
where $l$ is the curvature of this space. The index $z$ is called radial axis, and $i$ runs over $0,1,2,...,d$. The AdS space is a vacuum solution of the Einstein equation
\begin{eqnarray}
G^{\mu\nu}-\frac{1}{2}g^{\mu\nu}R+g^{\mu\nu}\Lambda=\kappa T^{\mu\nu}=0,
\end{eqnarray}
with negative cosmological constant $\Lambda=-d(d+1)/2l^{2}$.

\subsection{Killing equation and conformal invariance}

The Killing equation on the bulk AdS space approaches the conformal one at the boundary $z\rightarrow 0$. Let us start from infinitesimal transformation
\begin{eqnarray}
x^{\prime\mu}=x^{\mu}+\xi^{\mu}(x),
\end{eqnarray}
and then the new line element is given by
\begin{eqnarray}
ds^{\prime 2}=ds^{2}+\left(\nabla_{\mu}\xi_{\nu}+\nabla_{\nu}\xi_{\mu}\right)dx^{\mu}dx^{\nu},
\end{eqnarray}
where the covariant derivative is defined by
\begin{eqnarray}
\nabla_{\mu}\xi_{\nu}=\partial_{\mu}\xi_{\nu}-\Gamma^{\lambda}_{\;\mu\nu}\xi_{\lambda},
\end{eqnarray}
with the Christoffel symbol
\begin{eqnarray}
\Gamma^{\rho}_{\;\mu\nu}=\frac{1}{2}g^{\rho\tau}\left(\partial_{\mu}g_{\nu\tau}+\partial_{\nu}g_{\mu\tau}-\partial_{\tau}g_{\mu\nu}\right).
\end{eqnarray}
Here, the isometory $ds^{\prime 2}=ds^{2}$ yields the Killing equation $\nabla_{\mu}\xi_{\nu}+\nabla_{\nu}\xi_{\mu}=0$. We consider the following transformation
\begin{eqnarray}
\xi^{i}(z,x)&\rightarrow&\epsilon^{i}(x), \\
\xi^{z}(z,x)&\rightarrow&z\zeta(x),
\end{eqnarray}
where the boundary does not move after the transformation. Substituting these equations and the AdS metric into the Killing equation, we actually obtain the conformal Killing equation at the boundary
\begin{eqnarray}
\partial_{i}\epsilon_{j}+\partial_{j}\epsilon_{i}=\zeta\eta_{ij}.
\end{eqnarray}
Thus, the CFT lives on the boundary of the AdS space. Actually, it is possible for various models to prove
\begin{eqnarray}
\frac{\delta}{\delta\phi(x)}\frac{\delta}{\delta\phi(x^{\prime})}\left.\exp\left(-\frac{1}{2\kappa^{2}}I\right)\right|_{\phi=0}\propto\frac{1}{|x-x^{\prime}|^{\Delta}},
\end{eqnarray}
with the scaling dimension $\Delta$ of the CFT and $I$ denoting the classical Einstein-Hilbert action. This is so called Gubser-Krevanov-Polyakov-Witten relation~\cite{HM-Maldacena,HM-Gubser,HM-Witten,HM-Aharony}.

\subsection{Geometric interpretation of entanglement entropy}

An important guiding result to understand the physical background of the AdS/CFT correspondence is the so called Ryu-Takayanagi's formula for the entanglement entropy~\cite{HM-Ryu1,HM-Ryu2,HM-Nishioka}
\begin{eqnarray}
S=\frac{{\rm Area}(\gamma_{A})}{4G}, \label{HM-RT}
\end{eqnarray}
where $\gamma_{A}$ is the minimal surface whose boundary is given by the manifold $\partial\gamma_{A}=\partial A$, and $G$ is the Newton constant of gravity in the AdS space. This is general extension of the Bekenstein-Hawking entropy in Eq.~(\ref{HM-BH}). 

\begin{figure}
\centerline{\includegraphics[width=11cm]{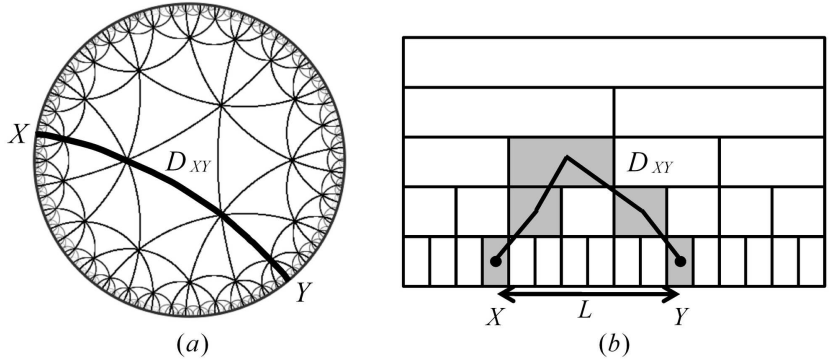}}
\caption{Two representations of 2D hyperbolic space and geodesic line: (a) Poincar$\acute{\rm e}$ disk model, (b) discrete model. In (b), $D_{XY}$ may not look like a geodesic line, but this is due to ambiguity of lattice discretization. The original AdS metric in Eq.~(\ref{HM-AdS}) can be transformed into $ds^{2}=(d\tau\log 2)^{2}+(2^{-\tau/l}dx)^{2}$. This form is comparable to Fig.~(b). }\label{HM-fig3}
\end{figure}

For $d=1$, ${\rm Area}(\gamma_{A})$ is the geodesic distance $D_{XY}$ between $X=(-L/2,a)$ and $Y=(L/2,a)$ for $a\rightarrow 0$ as shown in Fig.~\ref{HM-fig3}. These points are located on the boundary of the AdS space. Let us derive an explicit formula of $D_{XY}$, since this quantity plays a crucial role on the later discussion. The geodesic equation is given by
\begin{eqnarray}
\frac{d^{2}x^{\rho}}{d\tau^{2}}+\Gamma^{\rho}_{\;\mu\nu}\frac{d x^{\mu}}{d\tau}\frac{d x^{\nu}}{d\tau}=0,
\end{eqnarray}
and the solution is the half-cycle
\begin{eqnarray}
\left(x^{1},x^{2}\right)=(x,z)=\frac{L}{2}\left(\cos\theta,\sin\theta\right),
\end{eqnarray}
where $d\theta/dt=\sin\theta$, $\epsilon\le\theta\le\pi-\epsilon$, and $(L/2)\sin\epsilon=a$.
By using the equation for the geodesic line, we calculate $D_{XY}$ as
\begin{eqnarray}
D_{XY}&=&2\int_{\epsilon}^{\pi/2}\frac{l}{z}d\theta\sqrt{\left(\partial_{\theta}z\right)^{2}+\left(\partial_{\theta}x\right)^{2}} \nonumber \\
&=&l\ln\left(\frac{1+\cos\epsilon}{1-\cos\epsilon}\right) \nonumber \\
&=&2l\ln\left(\frac{L+\sqrt{L^{2}-(2a)^{2}}}{2a}\right).
\end{eqnarray}
By combining this with the Brown-Henneaux central charge~\cite{HM-Henningson}
\begin{eqnarray}
c=\frac{3l}{2G},
\end{eqnarray}
we can derive the CFT result $S=(c/3)\ln(L/a)$ for $L\gg a$. This calculation can be extended to higher dimensional cases.

It is noted that physical meanings of curved spaces are also explored in statistical mechanics. Hyperbolic deformation of interaction and sine-squared deformation are two major approaches~\cite{HM-Ueda,HM-Hikihara,HM-Maruyama,HM-Katsura}.

\section{Tensor Networks and Extra Dimension}

\subsection{From Matrix Product to Tensor Product}

\begin{figure}
\centerline{\includegraphics[width=6cm]{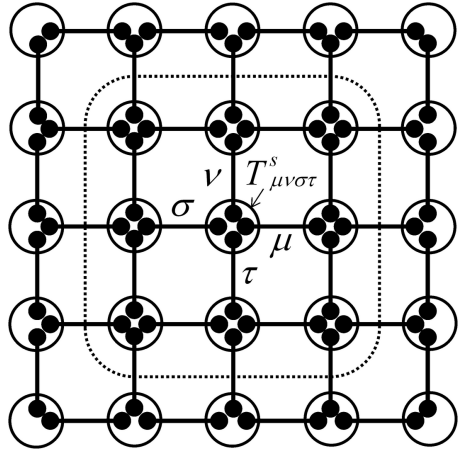}}
\caption{Tensor network structure on 2D square lattice. The tensor $T^{s}_{\mu\nu\sigma\tau}$ with linear dimension $\chi$ is defined on each lattice site. The quantum entanglement between neighboring sites is represented by the indices $\mu$, $\nu$, $\sigma$, and $\tau$.}\label{HM-fig4}
\end{figure}

As we have already mentioned, the matrix dimension $\chi$ of the MPS state represents how strong quantum entanglement appears. However, the MPS state is appropriate only in 1D gapped cases. Direct extention of this approach to higher dimensions is to construct tensor networks (or sometimes called tensor product state, TPS, and projected entangled-pair state, PEPS)~\cite{HM-Verstraete1,HM-Verstraete2,HM-Verstraete3,HM-Isacsson,HM-Murg1,HM-Murg2}. We show a schematic viewgraph of TPS in Fig.~\ref{HM-fig4}. We define a tensor, $T_{\mu\nu\lambda\cdots}^{s_{j}}$, on each site $j$. The number of the indices, $\mu,\nu,\lambda,\cdots$, of each tensor corresponds to that of connected bonds. If each bond is maximally entangled, the state of the bond can be represented by $\alpha$ and its ancila $\bar{\alpha}$
\begin{eqnarray}
\left|\phi\right>=\sum_{\alpha=1}^{\chi}\frac{1}{\sqrt{\chi}}\left|\alpha\bar{\alpha}\right>.
\end{eqnarray}
Then, the entropy for each bond is given by $S_{bond}=-tr\left(\rho\ln\rho\right)=\log\chi$, where the trace is taken over the ancila degree of freedom. The entanglement entropy of the subsystem marked by a dotted line in Fig.~\ref{HM-fig4} is given by
\begin{eqnarray}
S=NS_{bond}\sim \left(\frac{L}{a}\right)^{d-1}\log\chi,
\end{eqnarray}
where $N$ is the number of the entangled bonds, $L$ is linear size, and $a$ is lattice constant. This relation is nothing but the area law scaling. Thus, we know that the TPS state is suitable for $d$-dimensional gapped cases. On ther other hand, $\chi\sim L$ in critical cases, and further improvement of the network structure may be required. This is the purpose of the next section where tree tensor and MERA networks are introduced.

\subsection{Tree Tensor Networks and Multiscale Entanglement Renormalization Ansatz: Hierarchical Tensor Network}

\begin{figure}
\centerline{\includegraphics[width=8cm]{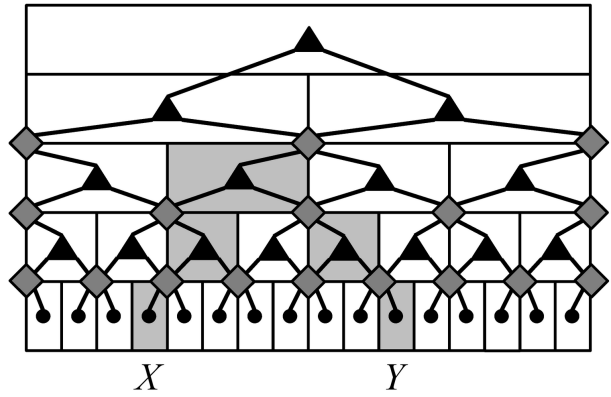}}
\caption{Schematic MERA network. Triangles and diamonds are isometory tensors and disentangler tensors, respectively. Each bond connecting a triangle with a diamond represents tensor product. For comparison, Fig.~\ref{HM-fig3} is combined with the network.}\label{HM-fig5}
\end{figure}

\begin{figure}
\centerline{\includegraphics[width=8cm]{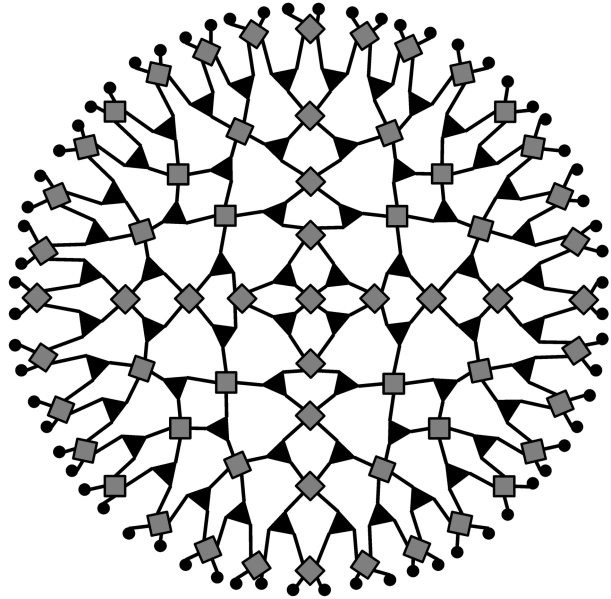}}
\caption{Deformation of MERA network in Fig.~\ref{HM-fig5}.}\label{HM-fig6}
\end{figure}

Constructing efficient variational optimization methods of MPS and TPS is of practical importance in condenced matter physics. An advantage of these methods is to keep the area-law scaling of the entropy, and thus is suitable to gapped $d$-dimensional systems. The application of tensor networks to critical systems leads to the tree tensor network (TTN) and the multiscale entanglement renormalization ansatz (MERA)~\cite{HM-Vidal2,HM-Vidal3,HM-Evenbly1,HM-Evenbly2,HM-Pfeifer,HM-Corboz,HM-Evenbly3,HM-Silvi}. They are spatially $(d+1)$-dimensional hierarchical networks that are also compatible with real-space renormalization group. In particular, the MERA network forms discrete AdS space as shown in Figs.~\ref{HM-fig5} and \ref{HM-fig6}~\cite{HM-Swingle2,HM-Evenbly4}. Although strict correspondence between AdS/CFT duality and MERA is not constructed in the present stage, they are very similar. When we express a critical system by using TPS, we need large tensor dimension $\chi\sim L$. On the othere hand, each tensor in the hierarchical network of MERA has relatively small dimension. Thus, we may say that MERA constructs more classical-like states in the holographic space.

\section{Compactification and Entropy}

\subsection{Bulk/edge Correspondence and Compactification}

\begin{figure}
\centerline{\includegraphics[width=8cm]{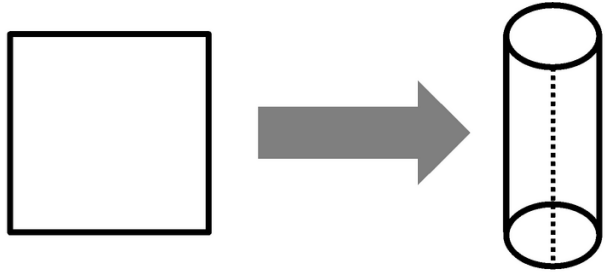}}
\caption{Compactification of open 2D sheet into 1D system}\label{HM-fig7}
\end{figure}

AdS/CFT is a kind of bulk/edge correspondence on non-compact manifold. Compactification of non-compact space is an alternative way to find correspondence between physically different systems. The comparison between these ways may be useful for further examination of the entanglement entropy. The concept of the compactification is schematically shown in Fig.~\ref{HM-fig7} where an open 2D sheet is rolled up. When our spatial resolution is worse than the compactification radius, this is effectively a 1D system. Unfortunately, we can not directly define the entanglement entropy in the classical space, since entanglement represents quantum correlation. However, according to the Ryu-Takayanagi's formula, a geometric quantity would correspond to the entropy. Then, the problem is to find geometrical meaning of the entanglement entropy in the non-compact manifold before compactification. In the following subsections, some historical review and my recent work closely related to this aspect are presented.

\subsection{Field theory with Extra Dimension}

Let us look at compactification of a massless scalar field model. The metric is given by the flat 4D Minkowski one ($i=0,1,2,3$) plus additional component $dx^{4}=a d\theta$
\begin{eqnarray}
g_{\mu\nu}=\left(
\begin{array}{cc}
\eta_{ij}&0 \\
0&a^{2}
\end{array}
\right),
\end{eqnarray}
where the radial direction is represented by $\theta$, and $a$ is its radius. We assume a cylindrical boundary condition for the $\theta$ direction, and the others are not. Then, we have $0\le\theta\le 2\pi$. The scalar field $\phi(x,\theta)$ is discretized as
\begin{eqnarray}
\phi(x,\theta)=\frac{1}{\sqrt{2\pi}}\sum_{n}\phi_{n}(x)e^{in\theta},
\end{eqnarray}
where $\phi_{-n}=\phi_{n}^{\ast}$. Substituting it to the action, the action after integrating over $\theta$ degree of freedom is given by
\begin{eqnarray}
I&=&\frac{1}{2\pi}\int d^{d}x\int_{0}^{2\pi}d\theta\sqrt{-g}\left(-\frac{1}{2}g^{\mu\nu}\partial_{\mu}\phi\partial_{\nu}\phi\right) \\
&=&a\int d^{d}x\left[ -\frac{1}{2}\partial_{i}\phi_{0}\partial^{i}\phi_{0}-\sum_{n\ge 1}\left\{\partial_{i}\phi_{n}^{\ast}\partial^{i}\phi_{n}+\left(\frac{n}{a}\right)^{2}\phi_{n}^{\ast}\phi_{n}\right\} \right].
\end{eqnarray}
Therefore, in addition to the zero mass mode $\phi_{0}$, we obtain the massive modes $\phi_{n}$ with mass $M_{n}=n/a$. They are called Kaluza-Klein modes. In the classical field theory which aims dimensional reduction, $a$ should be taken to be small enough. However, in the following, we are interested in cases of various $a$ values. For large $a$ values, these modes tend to be gapless. Thus, this indicates that the enough internal degree of freedom is necessary to describe low-energy excitations. This feature can also be seen in the tensor network formulation of quantum states.

\subsection{VBS/CFT correspondence}

Recently, it has been shown that the entanglement spectra of the VBS state on 2D lattices are closely related to a thermal density matrix of a holographic spin chain whose spectrum is reminiscent of that of the spin-$1/2$ Heisenberg chain~\cite{HM-Cirac,HM-Jie}. Here, we briefly touch on an idea of this correspondence based on the tensor product. It might be possible to have a different view based on open/close string duality. We start with the 2D VBS state on $M\times N$ lattice (vertical $\times$ horizontal) expressed by
\begin{eqnarray}
\left|\psi\right>=\sum_{I}c_{I}\left|I_{1}I_{2}\cdots I_{N}\right>,
\end{eqnarray}
where $I_{n}=(i_{1,n},i_{2,n},...,i_{M,n})$. Let us consider a cylindrical boundary condition that the vertical direction is rolled up, while the horizontal axis remains open. Then, we can introduce the coefficient $c_{I}$ defined by
\begin{eqnarray}
c_{I}=\sum_{\Lambda}L_{\Lambda_{1}}^{I_{1}}B_{\Lambda_{1}\Lambda_{2}}^{I_{2}}\cdots B_{\Lambda_{N-2}\Lambda_{N-1}}^{I_{N-1}}R_{\Lambda_{N-1}}^{I_{N}}, \label{HM-newMPS}
\end{eqnarray}
where $\Lambda_{n}=(\alpha_{1,n},\alpha_{2,n},...,\alpha_{M,n})$ and $\alpha_{j,n}=1,2,...,\chi$ with bond dimension $\chi$ of the original VBS. The periodic boundary condition along the vertical direction due to the cylinder form is thus expressed by
\begin{eqnarray}
B_{\Lambda_{n-1}\Lambda_{n}}^{I_{n}}=tr\prod_{j=1}^{M}A_{\alpha_{j,n-1}\alpha_{j,n}}^{i_{j,n}},
\end{eqnarray}
and its boundary terms $L_{\Lambda_{1}}^{I_{1}}$ and $R_{\Lambda_{N-1}}^{I_{N}}$. Clearly, Eq.~(\ref{HM-newMPS}) is a MPS form. When we devide this system into two parts, the state $\left|\psi\right>$ can be expressexd by the Schmidt decomposition
\begin{eqnarray}
\left|\psi\right>=\sum_{I_{a},I_{b}}\sum_{\Lambda}L_{\Lambda}^{I_{a}}R_{\Lambda}^{I_{b}}\left|I_{a},I_{b}\right>,
\end{eqnarray}
where $L^{I_{a}}=L^{I_{1}}B^{I_{2}}\cdots B^{I_{l}}$ and $R^{I_{b}}=B^{I_{l+1}}\cdots B^{I_{N-1}}R^{I_{N}}$. The dimension of $\Lambda$ is $\chi^{M}$. Originally, the VBS state has relatively small dimension $\chi$ due to the presence of the Haldane gap. However, $\chi^{M}$ becomes very large, leading to critical behavior. I think it's interesting to compare this idea with the compactification discussed in the previous subsection.

\subsection{Suzuki-Trotter Decomposition}

The readers expert for quantum Monte Carlo (QMC) simulation may be aware of the similarity between Suzuki-Trotter decomposition (STD) and compactification~\cite{HM-Suzuki}. The correspondence between classical and quantum systems are induced by the following STD
\begin{eqnarray}
e^{X+Y}=\lim_{M\rightarrow\infty}\left(e^{\frac{X}{M}}e^{\frac{Y}{M}}\right)^{M}
\end{eqnarray}
for non-commutative operators $X$ and $Y$. In some cases, the right hand side can be treated exactly.

It is well-known that the partition function of the transverse-field Ising chain can be mapped onto that of the 2D classical anisotropic Ising model. The transverse-field Ising Hamiltonian is given by
\begin{eqnarray}
H=H_{0}+H^{\prime}=-J\sum_{i=1}^{L}\sigma_{i}^{z}\sigma_{i+1}^{z}-\lambda\sum_{i=1}^{L}\sigma_{i}^{x}.
\end{eqnarray}
Let us derive the classical model with use of the STD. We introduce the partition function $Z={\rm Tr} e^{-\beta H}$, and decompose it into $M$ slices as
\begin{eqnarray}
Z&=& \sum_{\{\sigma_{1}\}}\left<\{\sigma_{1}\}\right| \left[ \exp\left(-\frac{\beta}{M}H_{0}\right)\exp\left(-\frac{\beta}{M}H^{\prime}\right) \right]^{M} \left|\{\sigma_{1}\}\right> \\
&=& \sum_{\{\sigma_{1}\},...,\{\sigma_{M}\}}\prod_{k=1}^{M}\left<\{\sigma_{k}\}\right| \exp\left(-\frac{\beta}{M}H_{0}\right)\exp\left(-\frac{\beta}{M}H^{\prime}\right) \left|\{\sigma_{k+1}\}\right> \\
&=& \sum_{\{\sigma_{1}\},...,\{\sigma_{M}\}}\prod_{k=1}^{M}\exp\left\{\frac{\beta}{M}J\sum_{i=1}^{L}\sigma_{i}^{k}\sigma_{i+1}^{k}\right\} \nonumber \\
&& \;\;\;\;\;\;\;\;\;\;\;\;\;\;\;\;\;\;\;\;\;\;\;\;\;\;\;\; \times\left<\{\sigma_{k}\}\right| \exp\left(-\frac{\beta}{M}H^{\prime}\right) \left|\{\sigma_{k+1}\}\right>.
\end{eqnarray}
Here, we have imposed the cylindrical boundary condition, $\{\sigma_{1}\}=\{\sigma_{M+1}\}$. Thus, we may say that the mathematical processes of the STD zoom up internal degrees of freedom on each site, which is analogous to the previous sections. For $\sigma,\sigma^{\prime}=\pm 1$, we have a relation
\begin{eqnarray}
\left<\sigma\right| e^{\frac{\beta}{M}\lambda\sigma^{x}}\left|\sigma^{\prime}\right> = A\exp\left\{-\frac{1}{2}\sigma\sigma^{\prime}\ln\left(\tanh\frac{\beta}{M}\lambda\right)\right\},
\end{eqnarray}
with $A=\sqrt{(1/2)\sinh(2\beta/M)\lambda}$. Thus, $Z$ is expressed as
\begin{eqnarray}
Z &=& A^{M}\prod_{k=1}^{M}\exp\left\{\frac{\beta}{M}J\sum_{i=1}^{L}\sigma_{i}^{k}\sigma_{i+1}^{k}\right\} \nonumber \\
&&\;\;\;\;\;\;\; \times\exp\left\{-\frac{1}{2}\sum_{i=1}^{L}\sigma_{i}^{k}\sigma_{i}^{k+1}\ln\left(\tanh\frac{\beta}{M}\lambda\right)\right\},
\end{eqnarray}
and finally the effective Hamiltonian is given by
\begin{eqnarray}
H_{\rm eff}=\sum_{i=1}^{L}\sum_{k=1}^{M}\left(J_{1}\sigma_{i}^{k}\sigma_{i+1}^{k}+J_{2}\sigma_{i}^{k}\sigma_{i}^{k+1}\right),
\end{eqnarray}
where $J_{1}=-J/M$ and $J_{2}=-(1/2\beta)\ln\left(\tanh(\beta\lambda/M)\right)$.

\subsection{Entropy scaling, quantum-classical correspondence, and hyperbolic geometry hidden in image processing based on SVD}

According to the previous discussion, let us finally examine what is a geometrical object corresponding to the entropy on the classical spin systems before compactification. That is a snapshot of a particular spin configuration at criticality.

We regard the orignal image data of the snapshot with $M\times N$ pixels as a matrix $\psi(x,y)$ ($1\le x\le M$, $1\le y\le N$). We assume that $\psi$ is real. Then, it is possible to introduce the density matrices
\begin{eqnarray}
\rho_{X}(x,x^{\prime})&=&\sum_{y}\psi(x,y)\psi(x^{\prime},y), \\
\rho_{Y}(y,y^{\prime})&=&\sum_{x}\psi(x,y)\psi(x,y^{\prime}).
\end{eqnarray}
Their $L$ non-zero eigenvalues are the same ($L={\rm min}(M,N)$). According to Eqs.~(\ref{HM-SVD}) and (\ref{HM-ABequal}), we can define the entropy. In order to intuitively understand scaling relations which this snapshot entropy satisfies, let us imagine a snapshot of the classical 2D isotropic Ising spin system at criticality. There, the spin configuration is fractal, since various sizes of the ordered clusters coexist due to the critical fructuation. This also means that various length scales are mixed. We can pick up many fractal subsystems from the original snapshot, but their patterns themselves are different with each other. Therefore, after correcting all of the possible patterns, they would cover the total information of the partition function. The information of thermal fructuation contained in the partition function is represented by the differen patterns. In that sense, only one snapshot is necessary at criticality. According to the STD, we expect that there exists a quantum 1D system that is transformed into 2D isotropic classical Ising model. Therefore, the snapshot entropy should obey the scaling relation equal to that of the critical 1D quantum systems. Actually, I have recently shown that this conjecture is at least correct in the 2D isotropic classical Ising model on $L\times L$ lattice~\cite{HM-Matsueda1}. I have obtained the following results at $T_{c}$
\begin{eqnarray}
S&=&-\sum_{l=1}^{L}\lambda_{l}\ln\lambda_{l}=\ln L - 2, \label{HM-image1} \\
S_{\chi}&=&-\sum_{l=1}^{\chi<L}\lambda_{l}\ln\lambda_{l}\sim\frac{1}{6}\ln\chi +\gamma, \label{HM-image2}
\end{eqnarray}
for sufficiently large $L$ and a positive $\gamma$ value. They are comparable to CFT and MPS results. Furthermore, when we visualize each layer of the SVD
\begin{eqnarray}
\psi^{(l)}(x,y)=U_{l}(x)\sqrt{\Lambda_{l}}V_{l}(y),
\end{eqnarray}
we know that $\psi^{(l)}(x,y)$ has its own length scale. The images of $\psi^{(l)}(x,y)$ with larger singular values contain more global spin structures. Extensive examinations suggest that the index $l$ corresponds to the radial direction of the discrete AdS space. The result means that the coarse graining of the image by tuncating small singular values correspondes to a flow from the boundary to the bulk on the AdS space. This would be related to the holographic renormalization group.

The entropy of more realistic images also obey a clear scaling relation analogous to the entanglement support of MPS~\cite{HM-Matsueda2}. This is surprizing, since we think it's impossible to determine 'the central charge' of the realistic images. We need more detailed analysis of this type of quantum-classical correspondence.

\section{Summary}

In this article, I have reviewed various scaling relations of the entanglement entropy and related topics. Since the entropy represents universality of the model considered, we can obtain more global viewpoints beyond the model itself. In particular, the scaling is a very powerful tool to look at quantum-classical correspondence. Finally, we should be carefull to a fact that the quality of the information is given by the wave function, not the entropy. In this respect, we need comprehensive study of the scaling analysis and numerical techniques of optimizing tensor networks.

\end{document}